\documentclass[usenatbib]{mn2e}

\usepackage{graphicx}
\usepackage{natbib}
\footnotesize
\newdimen\minuswidth    
\setbox0=\hbox{$-$}
\minuswidth=\wd0
\catcode`@=\active
\def@{\kern\minuswidth}
\newdimen\digitwidth    
\setbox0=\hbox{\rm0}
\digitwidth=\wd0
\catcode`!=\active
\def!{\kern\digitwidth}
\normalsize
\title[Pulsar parameter estimation]
{Pulsar timing analysis in the presence of correlated noise}
\author[W. Coles et al.]
{W. Coles,$^1$
G. Hobbs,$^2$
D. J. Champion,$^3$
R. N. Manchester,$^2$ 
J. P. W. Verbiest$^3$
\\
$^1$ Electrical and Computer Engineering, University of California at San Diego, La Jolla, California, U.S.A. \\
$^2$ CSIRO Astronomy and Space Science, Australia Telescope National Facility, P.O.~Box~76, Epping NSW~1710, Australia \\
$^3$ Max-Planck-Institut f\"ur Radioastronomie, Auf dem H\"ugel 69, 53121 Bonn, Germany  \\
}

\date{}

\begin{document}
\maketitle
\newcommand{\setthebls}{
}
\setthebls
\begin{abstract}
Pulsar timing observations are usually analysed with least-square-fitting procedures under the assumption that the timing residuals are uncorrelated (statistically ``white"). Pulsar observers are well aware that this assumption often breaks down and causes severe errors in estimating the parameters of the timing model and their uncertainties. Ad hoc methods for minimizing these errors have been developed, but we show that they are far from optimal. Compensation for temporal correlation can be done optimally if the covariance matrix of the residuals is known using a linear transformation that whitens both the residuals and the timing model. We adopt a transformation based on the Cholesky decomposition of the covariance matrix, but the transformation is not unique. We show how to estimate the covariance matrix with sufficient accuracy to optimize the pulsar timing analysis. We also show how to apply this procedure to estimate the spectrum of any time series with a steep red power-law spectrum, including those with irregular sampling and variable error bars, which are otherwise very difficult to analyse. 
\end{abstract}


\section{Introduction}

Pulsar timing provides a powerful tool for studying a wide range of phenomena ranging from basic physics, such as testing the general theory of relativity \cite[e.g.,][]{ksm+06}, through astronomy, for instance measuring pulsar positions and proper motions \cite[e.g.,][]{hllk05}, to looking for irregularities  in terrestrial time scales \cite[e.g.,][]{pt96,rod08}.  An overview of the pulsar timing technique is given in \cite{lk05} and details are provided in \cite{ehm06}.  In brief, a pulsar timing model is used to predict pulse times-of-arrival (ToAs) at the observatory.  These model predictions are compared with the measured ToAs and the differences are known as ``pulsar timing residuals''.  The timing model is subsequently improved using a least-squares-fitting procedure to minimize the timing residuals. 

\begin{figure}
\includegraphics[width=65mm,angle=-90]{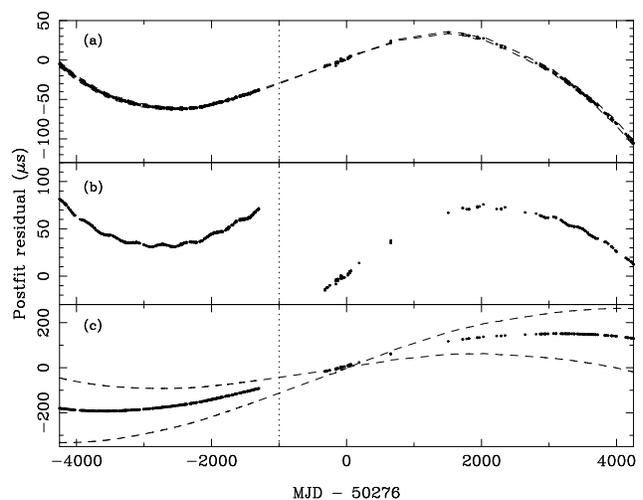}
\caption{Timing residuals for PSR~J1939$+$2134. Panel (a) shows the residuals from a standard pulsar timing model fit for the spin-frequency and its derivative. Panel (b) shows the residuals after also fitting for the position of the pulsar and a phase jump between Arecibo and Parkes observations using the simple weighted least squares method. Panel (c) shows the residuals after the same fitting as in panel (b) using the Cholesky method. The effect of changing $\nu$ by $\pm\sigma$ is illustrated in panels (a) and (c) by the dashed lines.}\label{fg:badexample}
\end{figure}

Since most of the parameters of the timing model are linear, at least for small perturbations, the fitting procedure is straightforward and the routines return an estimate of each of the parameters and the corresponding covariance matrix.  It has been the practice in pulsar timing to assume that the timing residuals are uncorrelated, although it is widely understood that this assumption is usually invalid.  Correlated timing residuals occur for many reasons, among them: inadequate calibration of the raw observations \citep[e.g.,][]{van06}; failure to correct for variations in the interstellar dispersion \citep[e.g.,][]{yhc+07}; and ``timing noise'' intrinsic to the pulsar.  Timing noise is still not fully understood, but usually refers to unexplained low-frequency features in the residuals \cite[e.g.,][]{lhk+10}. The main effects of neglecting correlation in the timing residuals are: (1) the parameters of the timing model are not estimated as accurately as possible and may have systematic biases; and (2) the uncertainties on the best-fit parameters are not correct. 

An extreme example of the problem can be seen in the 20-cm timing residuals for the millisecond pulsar J1939$+$2134 assembled by Verbiest et al. (2009)\nocite{vbc+09}. The first eight years of these observations were taken at the Arecibo Observatory \citep{ktr94} and the remaining observations were taken at the Parkes Observatory. There is an unknown phase discontinuity between the observations at the two observatories. An initial estimate of this ``jump'' was made by fitting the observations for only a year on either side of the jump with a model including the   pulse frequency $\nu$, its first derivative $\dot{\nu}$, and the jump. The fitting of $\nu$ and $\dot{\nu}$ removes a quadratic from the residuals, which in this case makes them almost white and allows for a reasonable initial estimate of the jump.
The residuals are displayed in Figure~\ref{fg:badexample} panel (a) after a weighted least squares (WLS) fit for $\nu$ and  $\dot{\nu}$ using the initial jump estimate. Fitting $\nu$ and $\dot{\nu}$ over the longer data span leaves an obvious cubic term in the residuals.
In panel (b) the residuals are shown after also fitting for the position  of the pulsar and the jump using WLS over the entire data span. The jump fitting over the entire data span is obviously catastrophic. An offset in position introduces an annual sine wave into the post-fit residuals.  Ideally the fitting process would minimize the power in the spectrum at 1~y$^{-1}$, but the high degree of correlation in the residuals introduces such a large error in the estimated position that it actually adds power at 1~y$^{-1}$. This shows as a distinct annual ripple in panel (b). 
Panel (c) shows the result of fitting the same parameters over the same data span as panel (b) using the new method that we describe in this paper. 
The new method, which we refer to as the Cholesky method, obviously fits the position and jump much better than panel (b) but there is an obvious trend in panel (c). This is because the estimate of $\nu$ in panel (c) is quite different from the one in panel (a). In panel (a) some of the red noise is absorbed into the $\nu$ estimate. The error in the trend is much larger than one would expect, due to the red noise. The effect of changing $\nu$ by $\pm\sigma$ is illustrated in panels (a) and (c) by the dashed lines. One can see that the trend in panel (c) is well within the uncertainty of the $\nu$ estimate. The estimation of $\nu$ and $\dot{\nu}$ will be discussed in more detail in section 5.3.

Another example of this problem is a comparison of the parallax of PSR~J0437$-$4715 as estimated from a timing analysis \citep{vbv+08}, with that estimated from recent Very Long Baseline Interferometric (VLBI) observations \citep{dvtb08}. The VLBI parallax is 6.396$\pm$0.054\,mas. The timing parallax estimated using a WLS fit was 6.65$\pm$0.07\,mas and we were able to duplicate this using the same observations. We obtain a value of 6.34$\pm$0.12\,mas using the Cholesky method. Clearly the Cholesky result is consistent with the VLBI result and the WLS method is not. It should be noted that Verbiest et al. (2008) \nocite{vbv+08} were doubtful of the formal error estimate for the WLS method and they re-estimated the error using a Monte Carlo simulation which increased the error from 0.07\,mas to 0.51\,mas. As the actual difference is 0.25\,mas this revised error estimate may have been conservative. It is quite common for pulsar observers to be suspicious of the error estimates obtained using WLS fits and to modify them in various ad hoc ways. We will show that using the Cholesky method eliminates this problem for all parameters of the timing model with time scales shorter than the observing span.  $\nu$ and $\dot{\nu}$ always have time scales comparable with the observing span and will be discussed separately in section 5.3.

Pulsar observers have often attempted to improve parameter estimates by removing some portion of the low-frequency timing noise, taking care not to remove the components that are needed to estimate the parameters of interest. The low frequencies have been removed by adding them to the timing model, either as a high order polynomial \citep[e.g.,][]{tacl99} or as a carefully chosen Fourier series (e.g., \textsc{fitwaves} in {\sc Tempo2}; Hobbs et al. 2004\nocite{hlk+04}). In either case the residuals are ``flattened'' or ``whitened'' and an adequate fit for position can be obtained (throughout this paper timing residuals that are uncorrelated are termed ``white'' and residuals that exhibit a steep low-frequency spectrum are termed ``red''). Neither method produces a good fit to a phase jump because the effect of a phase jump is not localised in frequency in the residual spectrum.

In this paper we describe a method of optimizing the least-squares fit by finding a linear transformation which whitens and normalizes the residuals. This transformation is then applied to both the observations and the timing model. The parameters can then be found by fitting the transformed timing model to the transformed observations using ordinary least-squares. The transformation can be
found exactly if the covariance matrix of the residuals is known, although it is not unique. This process is equivalent to the so-called ``generalized least-squares'' (GLS) solution, indeed it is how that solution was discovered. This method is not new, but it has not been applied to pulsar timing observations before. The GLS solution provides the best linear unbiased estimator of the parameters of the timing model and 
the best unbiased estimator of the covariance matrix of the estimated parameters. In most cases the covariance matrix of the residuals is not known and must be estimated from the observations. We have developed a procedure for estimating the covariance matrix which works well for the pulsars we have tested and we find that the parameter estimates are not unduly sensitive to errors in estimating the covariance matrix. It should be noted that the residuals formed by subtracting the timing model with the best fit parameters from the observations will not be white as can be seen in Figure~1 panel (c). The advantages of the Cholesky method over the polynomial or Fourier methods are: (1) it provides more accurate parameter estimates; (2) it provides a more accurate estimate of the covariance matrix of the parameters; (3) it does not require any ``adjustment'' to fit different parameters of the timing model.

In section 2 we outline the theory of linear least-squares fitting when the covariance matrix of the residuals is known. When the covariance is not known one must usually attempt a spectrum analysis of the residuals. In section 3 we discuss the problem of spectrum analysis of steep red random processes which are sampled irregularly and each sample has a different uncertainty.  In section 4 we describe our procedure for estimating the covariance matrix of the residuals by spectrum analysis and demonstrate it on two pulsars with different types of timing noise. In section 5 we show that the Cholesky method is consistently superior to the WLS, polynomial, and Fourier methods, using simulations to make the comparisons more precise. Finally in section 6 we use simulations to establish the sensitivity of the Cholesky method to errors in estimating the covariance matrix. The tools that we discuss have been incorporated into the \textsc{Tempo2} \citep{hem06} timing analysis package and are available on-line\footnote{http://www.atnf.csiro.au/research/pulsar/tempo2} for the community. A step-by-step tutorial on using the Cholesky method is also available on the \textsc{Tempo2} web site.

\section{Linear Least Squares Theory}


The least-squares formalism separates the observations into a deterministic component, the timing model, and a (zero mean) random component. The random component is referred to in the statistical literature as the ``error'' and in the pulsar community as ``post-fit timing residuals''. The least-squares is linear if the timing model is a linear function of the parameters which must be estimated. The formalism can be compactly described using matrix algebra as follows. Here $n$ timing residuals are modelled using a system of linear equations in $m < n$ parameters,
\begin{equation}
\vec{R} = {\bf M}\vec{P} + \vec{E}
\end{equation}
where $\vec{R}$ is an $n$-point column vector representing the pre-fit timing residuals, ${\bf M}$ is an $n \times m$ matrix describing the pulsar timing model, $\vec{P}$ is an $m$-point column vector containing the fitted parameters and $\vec{E}$ is an $n$-point column vector representing the post-fit residuals. 
If the post-fit residuals are independent with equal variance ($\sigma^2$) for all observations (homoscedastic), then one minimises the squared error. 

In vector notation the squared error is
\begin{equation}
\vec{E}^T \vec{E} =  (\vec{R} - {\bf M}\vec{P})^T(\vec{R} - {\bf M}\vec{P})
\end{equation}
and the solution, which is called ordinary least squares (OLS), is
\begin{equation}
\vec{P}_{\rm est} = ({\bf M}^T {\bf M})^{-1} {\bf M}^T \vec{R}.
\end{equation}
If the residuals have a gaussian distribution this is also ``maximum likelihood solution''.
The covariance matrix of the parameters is given by
\begin{equation}
cov(\vec{P}_{\rm est}) = \langle \vec{P}_{\rm est}\vec{P}_{\rm est}^T \rangle = \sigma^2 ({\bf M}^T{\bf M})^{-1}.
\end{equation}
Here the angle brackets denote a statistical expectation.

The normalised squared error $\vec{E}^T \vec{E} / \sigma^2$ is a chi-squared random variable with $n-m$  degrees of freedom. It is often referred to as $\chi^2$ and used as part of a ``goodness-of-fit test." It is the normal practice to scale $cov(\vec{P}_{\rm est})$ by $\chi^2/(n-m)$ at least when $\chi^2 > (n-m)$. If $\chi^2 < (n-m)$ there may be a problem. Either the data may have been ``overfit'' or $\sigma^2$ may have been underestimated.
If $\sigma^2$ is unknown a priori, then one estimates it using
\begin{equation}
\sigma_{est}^2 = \vec{E}^T \vec{E}/(n-m).
\end{equation}

If the residuals are white, but each ToA has a different variance (heteroscedastic), then one minimises the weighted squared error (the WLS solution). This approach is widely used in pulsar timing analysis. For white residuals, the covariance matrix of the residuals is ${\bf V}$ a diagonal matrix with the variance of the samples on the main diagonal. In this form the weighted squared error is $\vec{E}^T {\bf V}^{-1} \vec{E}$ and the solution is
\begin{equation}
\vec{P}_{\rm est} = ({\bf M}^T {\bf V}^{-1}{\bf M})^{-1} {\bf M}^T {\bf V}^{-1}\vec{R}.
\end{equation}
The same result can be obtained by normalizing both the data and the model by multiplying each by ${\bf V}^{-0.5}$. That is 
$\vec{R_N} = {\bf V}^{-0.5} \vec{R}$ and ${\bf M_N} = {\bf V}^{-0.5}{\bf M}$. Then the errors will all have unit variance and
the solution is 
\begin{equation}
\vec{P}_{\rm est} = ({\bf M_N}^T {\bf M_N})^{-1} {\bf M_N}^T \vec{R_N}.
\end{equation}
Thus the WLS solution becomes an OLS problem in the normalized variables. If the normalization is correct the minimum squared error $\vec{E_N}^T \vec{E_N}$ will have a $\chi^2$ distribution with $n-m$ degrees of freedom. The covariance matrix of the parameters is $(\chi^2/(n-m))({\bf M_N}^T{\bf M_N})^{-1}$. It must be scaled by the measured $\chi^2$ to allow for errors in the normalization.

The OLS  and WLS solutions have been known since the work of Gauss and Legendre at the beginning of the 19th century. The WLS analysis can be extended to the case of correlated residuals if the covariance matrix of the residuals ${\bf C} = \langle \vec{E} \vec{E}^T\rangle$ is known.
In this case one minimises $\vec{E}^T {\bf C}^{-1}\vec{E}$ and the solution (Aitken, 1935)\nocite{ait35} is often referred to as generalised least squares (GLS).
\begin{equation}\label{eqn:optimal}
\vec{P}_{\rm est} = ({\bf M}^T{\bf C}^{-1} {\bf M})^{-1} {\bf M}^T{\bf C}^{-1} \vec{R}.
\end{equation}
The GLS solution was derived, and is best described, as a normalizing and whitening process.  ${\bf C}$ is Hermitian positive semi-definite and so can be factored into ${\bf C} = {\bf U}{\bf U}^T$ using a Cholesky lower triangle factorisation. The matrix ${\bf U^{-1}}$ can be used as a normalizing and whitening transformation (but is not unique, the Mahalanobis transformation can also be used). Defining $\vec{E}_{\rm w} = {\bf U}^{-1}\vec{E}$, $\vec{R}_{\rm w} = {\bf U}^{-1}\vec{R}$ and ${\bf M}_{\rm w} = {\bf U}^{-1}{\bf M}$, we find that $\rm{cov}(\vec{E}_{\rm w}) = {\bf I}$, the identity matrix. The transformed residuals $\vec{E}_{\rm w}$ are therefore both white and normalized to unit variance. The GLS solution is now an OLS problem in the transformed variables, 
and Equation~(8) can be rewritten as
\begin{equation}
\vec{P}_{\rm est} = ({\bf M}^T_{\rm w} {\bf M}_{\rm w})^{-1} {\bf M}^T_{\rm w} \vec{R}_{\rm w}.
\end{equation}
Again if the normalization is correct $\chi^2 = \vec{E_{\rm w}}^T \vec{E_{\rm w}}$ is a $\chi^2$ random variable with $n-m$ degrees of freedom and the covariance matrix of the parameters is
$(\chi^2/(n-m))({\bf M_{\rm w}}^T{\bf M_{\rm w}})^{-1}$.

In summary, all linear least-squares problems can be reduced to ordinary least-squares by a suitable transformation. However, to
find this transformation we must know the covariance matrix of the residuals at least within a constant multiplier. We will defer until section 4 the discussion of how to estimate the covariance matrix from the observations. The least squares solution, when properly transformed, provides minimum variance, linear unbiased estimators of the model parameters and their uncertainties.
It should be noted that inverting matrices of the form $({\bf M}^T {\bf M})$ directly is not computationally efficient and other algorithms should be employed. \textsc{Tempo2} uses a singular value decomposition.

Equations (8) and (9) are the same, so it does not matter which form one uses. However one must, in most cases, first estimate the covariance matrix and confirm that it is correct. This requires the Cholesky (or equivalent) transformation. So the transformation must be found and applied in any case. In practice we find it efficient to use an OLS solver for all cases so we first form ${\bf M}_{\rm w}$ and $\vec{R}_{\rm w}$ then use equation (9).

\section{Least Squares Spectral Analysis}

Spectral analysis is intimately connected to the analysis of pulsar timing observations in two ways. First, the timing model includes a number of parameters which control the amplitude and phase of sine waves, so fitting for those parameters amounts to a spectral analysis. Second, optimal least squares fitting requires estimation of the covariance matrix and this will require some form of spectral analysis. Here we provide a very brief discussion of the aspects of spectral analysis important to the analysis of pulsar timing observations. We will not attempt to provide a complete mathematical description because there are many books devoted to the subject (e.g., Blackman and Tukey, 1959, Jenkins and Watts, 1969; Manolakis, Ingle and Kogon, 2005)\nocite{bt59}\nocite{jw69}\nocite{mik05}.

Spectral analysis of a series of regularly spaced observations of length T is normally performed via the discrete Fourier transform (DFT). The squared magnitude of the DFT, properly scaled, known as the periodogram $P(f)$, is often used as the spectral estimator. Here the sample interval is $\delta$ and we normalize $P(f)$ to have the units of a power spectral density.

\begin{equation}
P({\rm f} = k/T) = (T/n^2) \left| \sum_{l=0}^{n-1} r(t= l\delta) {\rm e}^{-j 2 \pi k l \delta /T} \right| ^2
\end{equation}

However, $P(f)$ will be biased if the spectrum being estimated is not white. The bias is caused by the finite length of data. The data are effectively multiplied by a time window $w(t)$, which is often rectangular (i.e., it is unity during the observations and zero otherwise). Thus the Fourier transform of $r(t)w(t)$ is the convolution of $R(f)$ with $W(f)$. The measured power spectrum $P(f)$ is the convolution of the spectral density $S(f)$ with $|W(f)|^2$. For a rectangular window of length T the spectral window is given by
\begin{equation}
|W(f)|^2 = |\sin (\pi f T)/(\pi f T) |^2.
\end{equation}
This window is most widely used because it provides the highest frequency resolution. This is particularly important in pulsar timing analysis. The sidelobes of this window fall off as $f^{-2}$. This makes power law spectra which fall faster than $f^{-2}$ heavily biased because window sidelobes will dominate all frequencies higher than $f=1/T$. Steeper spectra must be analysed with ``prewhitening'' and ``postdarkening'' to minimize spectral leakage (e.g., Jenkins and Watts, 1969). One applies a linear pre-whitening filter which is implemented in the time domain. The purpose of this filter is to make the spectrum close enough to white that leakage is insignificant. For example, if $x(k)$ is the input and $y(k)$ is the output, a first difference filter is $y(k) = x(k) - x(k-1)$. The effect of this filter is to multiply the Fourier transform of the input by the transfer function $H(f) = 2 \sin(\pi f \delta)$. This multiplies the power spectrum by $|H(f)|^2 = (2 \sin( f \delta))^2$, which has an $f^2$ behavior at low frequencies. Thus it will whiten a power law spectrum with exponent $-$2. The whitened data can be analyzed with little spectral leakage if its spectral exponent is between -0.5 and -3.5. We then correct the estimated spectrum of the whitened data by dividing it by $|H(f)|^2$. This is called post-darkening.

For steeper spectra a second difference (two applications of the first difference) will be required. The transfer function of the second difference is $H(f)^2$. Unfortunately spectra often are power law at low frequencies, descending into white noise at high frequencies. The white noise will be transformed to $f^4$ behavior and spectral leakage can occur backwards from high frequencies to low frequencies. To prevent this one must also filter the original observations with a low-pass filter, implemented in the time domain, which has a transfer function of the form
$L(f) = 1/(1 + (f/f_c )^2)$. Here the corner frequency $f_c$ is chosen to be near the frequency at which the power law component reaches the white noise level.
The residuals after second differencing and low pass filtering are then spectrum analyzed. The post-darkening is done by dividing the estimated spectrum of the filtered data by $|L(f)|^2 |H(f)|^4$.

Since we do not know the spectral exponent a priori, the analysis must be done in stages. A first estimate can be done with a periodogram. If it appears to be limited by leakage then another analysis must be done using first difference pre-whitening and post-darkening. If it still appears to be limited by leakage one must do another analysis using second differencing and low pass filtering. In some cases a more complex pre-whitening filter may be required, but we have not seen any such cases in pulsar timing analyses.

We can apply pre-whitening and post-darkening, as described above, only to regularly sampled measurements because the temporal filtering operations requires equally spaced data. Fortunately this is possible for pulsar timing applications because the low frequency spectra are often power-law and the high frequencies are white. We can use a low pass filter to separate the low and high frequencies. The low frequencies are heavily over-sampled and can be interpolated onto a regular grid without much bias. The high frequencies cannot be interpolated, but as they are white they will not suffer from spectral leakage and we can use existing least-squares methods. Thus we can assemble a composite spectrum by splitting the spectrum with a low pass filter and analyzing the two halves with different methods
 
The DFT is equivalent to a least squares fitting of complex exponentials, but since the complex exponentials are orthogonal in the regularly sampled DFT, they can be fit either individually or simultaneously and the result will be the same. If the sample spacing is not regular the complex exponentials are not orthogonal. It is not possible to fit them all simultaneously because they are highly covariant. However we can obtain unbiased estimates of the variance at each frequency using the Lomb-Scargle L-S (Scargle, 1982)\nocite{sca82} or Z-K (Zechmeister and Kurster, 2009)\nocite{zk09} methods. This is sufficient for our purposes. These methods are essentially least-squares fits of a sine-cosine pair independently at each frequency, and scaled so the expected value of each component is the same if the input time series is white noise. The LS method is unweighted and the Z-K method is weighted. These methods also have window functions $W(f)$ but their windows have much higher sidelobes than does $\sin (\pi f T)/(\pi f T)$.

If the covariance matrix is known, we can estimate the spectrum of irregularly sampled data using the Cholesky least squares method. Of course if the covariance matrix were known exactly we would know the spectrum and would not need to do a spectrum analysis. Fortunately we do not need to know the covariance matrix exactly, for the same reason we do not need to do the prewhitening exactly, it is only necessary to whiten the spectrum enough to eliminate leakage. If the window sidelobes are lower than the main lobe by a factor of 10, then one can tolerate local variation in the spectrum of approximately a factor of 10. The Cholesky method essentially extends the L-S or Z-K methods by fitting a sine-cosine pair at each frequency, but first whitening both the data and the model (the sine-cosine pair) with the Cholesky transformation.

We show some examples of least-squares spectrum analysis in Figure~2. Here we simulated a steep red spectrum with white noise under two sampling regimes. This was done independently of {\sc tempo2}. We simulated the red component in the Fourier transform domain creating a daily sampled time series 100 times longer than the actual data span. We interpolated this to the desired sample times using a constrained cubic spline interpolator. Then we added the sampling noise. This provided 100 realisations of the process.
In panel (a) we used regular sampling with equal errors and in panel (b) we used the actual sampling of PSR~J1713+0747 and the corresponding errors from the Verbiest et al. (2009) data set. In both cases we fitted and removed a quadratic, as would always be done with pulsar timing observations. The spectra of 100 realisations of the random process have been averaged to reduce the estimation errors. In each case we have shown the actual spectrum as a short dashed line. 
The Z-K spectral estimates are shown as long dashed lines. In the upper panel the sampling is regular and the weights are constant so the DFT estimate, the Z-K estimate, and the L-S estimate are the same. One can see that the Z-K estimate suffers from very serious spectral leakage and the quadratic removal has reduced the first spectral estimate by a factor of about 10. In the lower panel the Z-K estimate shows even stronger spectral leakage, but quadratic removal has not reduced the first spectral estimate as severely as in the upper panel.
In both panels the Cholesky spectral estimates are shown as solid lines. In the upper panel this is an excellent match to the actual spectrum. In the lower panel the shape is correct but the Cholesky estimate is about a factor of two higher than the actual spectrum. Here the known spectrum was used to find the covariance matrix and the Cholesky transformation. Of course this is not possible with observations. We will discuss the iterative approach we use for observations in the following section.

One can see that in both panels the noise level at f = 1~${\rm y}^{-1}$ with the Cholesky estimate is much lower than with the Z-K estimate. The uncertainties on an estimate of position or proper motion would be correspondingly lower. In panel (a) one can see the window sidelobes of the WLS estimator fall like $f^{-2}$ as expected. In panel (b) the sidelobes do not continue to fall with increasing $f$. This is because they are dominated by the effects of irregular spacing and variable errors. This window is equivalent to the ``dirty beam'' in aperture synthesis with a sparsely sampled aperture. In panel (a) the excellent agreement between the mean Cholesky spectrum and the actual spectrum shows that the Cholesky method is effective in eliminating leakage. We believe that this is the first demonstration of the Cholesky transformation for this purpose. In panel (b) the mean Cholesky spectrum appears to be biased high by a factor of approximately 2. We think that this is due to the integrated effect of the sidelobes of the dirty beam. We have not (yet) found an analytical way to remove this bias, but we believe that it could be calibrated out using simulations. Fortunately it is not necessary to remove it for least-squares fitting of a timing model because the $\chi^2$ factor will absorb any constant factor, provided that the shape of the spectrum is correct.

We have also computed the covariance matrices of the Cholesky spectral estimates and find that they are essentially independent, whereas those made using WLS are highly dependent on the first spectral estimate. This is a very important point because it makes optimum detection of a power-law process
possible in the frequency domain. In this case the normal procedure would be prewhitening followed by
low-pass filtering and then estimation of the variance and possibly cross-covariances. It is equivalent to Wiener filtering. If one Fourier transforms the residuals then spectra and possibly cross-spectra
can be estimated. Each spectral estimate for which the signal is greater than the noise will provide two degrees of freedom. If the spectral estimates can be made independent then a weighted sum of the
spectral (and cross-spectral) estimates is exactly equivalent to the optimal Wiener filter.

\begin{figure}
\includegraphics[width=75mm,angle=0]{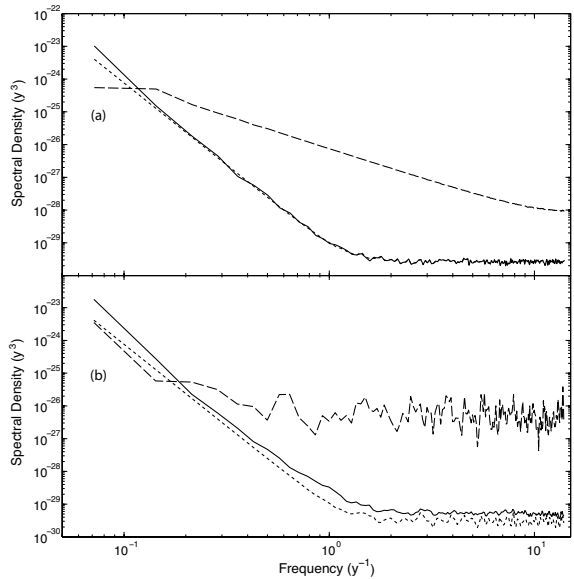}
\caption{Simulated power spectra. Each spectrum is the average of 100 realisations. The short dashed lines show the theoretical spectra. The long dashed lines are the WLS spectral estimates and the solid lines are the Cholesky spectral estimates. Panel (a) shows the spectra for regular sampling and equal weights. Panel (b) shows the spectra for irregular sampling and variable weights.}\label{fg:spectra}
\end{figure}

\section{Covariance Matrix Estimation}

The covariance matrix of $n$ observations contains $n(n-1)/2$ terms so it is not feasible to estimate the entire matrix from the $n$ observations without a simplifying statistical model. Here we assume that the timing residuals can be modeled as the sum of two random processes: the correlated timing noise $x(t)$, and the uncorrelated measurement error $e(t)$. We assume that $x(t)$ can be modeled as a wide-sense stationary process with a covariance function $c(\tau) = \langle x(t)x(t+\tau) \rangle$. Its power spectrum $P(f)$ is the Fourier transform of $c(\tau)$. The covariance matrix for $x(t)$ is ${\bf C}$ where each element ${\bf C_{ij}} = c(\tau_{ij})$  and $\tau_{ij} = | t_i - t_j |$. The timing residuals are sampled irregularly and $e(t)$ is a point process consisting of uncorrelated measurement errors which have different variance at each sample time. Its covariance matrix is a diagonal matrix ${\bf V}$, where the components of the diagonal are the measurement variances for each sample. The overall covariance matrix is simply the sum of ${\bf V}$ and ${\bf C}$.

   There must be cases where the assumption that the timing noise is wide-sense stationary breaks down since the causes of timing noise are incompletely understood. This would be a very interesting area of further research. There is also a subtle but important difference between the statistics of the pre-fit and post-fit residuals. Even if the pre-fit residuals are wide-sense stationary the post-fit residuals will not be. Since we actually have to estimate the covariance matrix of the post-fit residuals, there is always some break-down in this assumption. Fortunately the effects of such break-downs can be estimated through simulations as will be discussed in sections 5 and 6.

The observed TOAs are always accompanied by uncertainty estimates $\sigma_e (i)$, but these are generally insufficient to fully characterise the white noise. The $\sigma_e (i)$ are derived from the process of fitting a scaled and shifted template to the mean pulse shape. They will correctly describe the timing uncertainty caused by additive white gaussian noise, such as radiometer noise, but there are other potential sources of noise which can cause an uncertainty in the apparent pulse time of arrival. 
Observers often find that the ``scatter'' in the residuals appears to be larger than the error bars by a factor which is typically about two. This factor can be quite different for different receiving systems.   The ``fixData" plugin to \textsc{Tempo2} can be used to account for this unexplained scatter by increasing the $\sigma_e(i)$ estimates.  The plugin uses the mean structure function on short time lags as an estimate of the high frequency rms residual, and scales $\sigma_e (i)$ so it becomes independent of the receiving system. When the errors have been properly scaled computation of V is straightforward.

The estimation of the covariance matrix {\bf C} of red timing noise is not trivial, but it is easy to check that the transformed residuals are actually white and normalized. This is best done by estimating the spectrum using the L-S algorithm which is unbiased for such a random process. So we use an iterative approach:  estimating the spectrum; fitting it with a parametric model; using that model to find the covariance matrix; finding the Cholesky transformation; transforming the data; and estimating the spectrum of the transformed data with the L-S algorithm. This spectrum should be white. If it is much flatter than the original, as it always has been during our testing, we use this transformation to estimate the spectrum of the data using the Cholesky least squares method. From this estimate we obtain improved models of the spectrum; the covariance matrix; the transformation matrix, and ultimately a further improved estimate of the spectrum. As noted earlier, it is not essential that the whitened data have a perfectly flat spectrum, only that it be sufficiently flat that spectral leakage is not dominant. The first spectral estimate is obtained in two different ways, depending on whether the red noise in the residuals is ``weak'' or ``strong''. The entire process can be tested by simulation, including the effect of transforming the data with the wrong transformation matrix.

\subsection{Weak Red Noise}

The uncertainty in any estimated covariance function, such as $\tilde{c}(\tau)$ for the timing noise,
has the form $\delta \tilde{c} \approx (\tau_0/T_{obs})^{0.5} c(0)$ where $c(\tau_0) = c(0)/2$ (Jenkins and Watts 1969).
By comparison the uncertainty in estimating a white noise variance $\tilde{\sigma}^2$ has the form
$\delta \tilde{\sigma}^2 = \sigma^2/n^{0.5}$.
Since the time scale $\tau_0$ for red timing noise can approach $T_{obs}$ the error in
the covariance matrix can be large.
However in cases where $\tau_0 \ll T_{obs}$ and the variance of the timing noise is
comparable with the variance of the white measurement error, one can estimate $c(\tau)$
directly with adequate accuracy. One simply subtracts the mean and sums the pairwise products,  
averaging them into suitable ``$\tau$ bins'' weighted by their estimated uncertainties.
For 19 of the 20 millisecond pulsars in the Parkes Pulsar Timing Array this approach gives
a good estimate of $c(\tau)$. The exception, PSR~J1939+2134, which has $\tau_0 \approx T_{obs}$,
is shown in Figure 1. Irregular sampling is not a problem for any of the other pulsars in the
Verbiest et al (2009) data set for which $\tau_0 \ll T_{obs}$, because the distribution of time
differences $\tau$ between sample pairs, is much more uniform than the distribution of sample times.

In such cases we have found that an exponential model $c(\tau) =  
c(0) \exp ( - | \tau / \tau_0| )$, providing an amplitude and a time scale, is sufficient.
The exponential model is perhaps the simplest physical model, a single  
time-constant, and it has smooth behavior in the frequency domain,  
i.e., $P_x (f) = 2 c(0) \tau_0 / (1 + (2 \pi f \tau_0 )^2)$.
We fit this model to $c(\tau)$ to obtain $c(0)$ and $\tau_0$. The  
residuals for the millisecond pulsar J1045$-$4509  \citep{vbc+09}, which provide  
an example of this, are shown in Figure 3(a). One can see that there  
is low frequency noise present, but it is not dominant. The estimated  
covariance $c(\tau)$ is shown in Figure 3(b). Here the variance of the residuals 
is marked with an `o' symbol on the y-axis for comparison with the low frequency
variance c(0).
The best fit exponential model is shown as a solid line. The time  
scale is clearly much less than the data duration, and the variance in  
the white noise is greater than that in the red noise. The OLS  
spectrum of the residuals whitened using the Cholesky method is shown in Figure 3(c). It is  
white within the estimation errors.

\begin{figure}
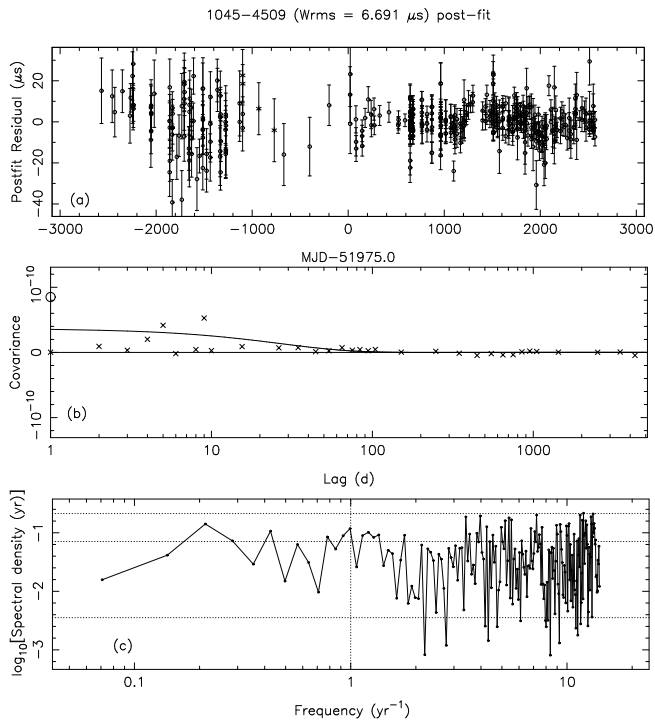

\includegraphics[width=34mm,angle=-90]{plot3a.ps}
\includegraphics[width=29.5mm,angle=-90]{plot3b.ps}
\includegraphics[width=30.5mm,angle=-90]{plot3c.ps}
\caption{Analysis of PSR~J1045$-$4509. Panel (a) shows the  
residuals from a standard pulsar timing model fit for $\nu$  
and $\dot{\nu}$. Panel (b) shows the estimated covariance function of  
the residuals. The variance, which includes the white noise, is marked 
with a  circle on the y axis.  
The best-fit exponential model is drawn as the solid line through the 
points which are the covariance measurements averaged into logarithmic bins. 
Panel (c)  
shows the OLS power spectrum of the residuals whitened and  
normalised with the Cholesky transformation. The horizontal dotted lines are the  
expected mean and 95\% confidence intervals for normalised white noise  
with this sampling.}
\label{fg:1045-4509}
\end{figure}

\subsection{Strong Red Noise}

The direct estimation of $c(\tau)$ breaks down when the low-frequency variance in the timing noise is dominant. In this case the residuals show a slow variation that substantially exceeds the error bars, as shown in Figures 1(a) and 4(a). In most such cases the power spectrum has the form $P(f) = A f^{-\alpha}$ where $\alpha > 2$. It is hard to estimate $c(\tau)$ because there are few degrees of freedom (i.e., $\tau_0 \approx T_{obs}$). However it is usually possible to make a power-law model of the spectrum and to specify the amplitude of that power law with reasonable accuracy. 
This is because each spectral estimate for which $P(f)$ is greater than the white noise, can provide two degree of freedom. As noted earlier, one must use a spectral estimator which provides independent estimates of $P(f)$.

Steep red processes require whitening, but we do not yet have the covariance matrix so we have to obtain a spectral estimate iteratively. We start by low-pass filtering the residuals to separate the red and white components. The resulting red component can be interpolated on to a regular grid without much distortion because it is quite smooth after the low-pass filtering. We can then pre-whiten it with a first difference process, compute the $|{\rm DFT}|^2$, and post-darken the result. This generally gives an adequate ``first guess'' at the power spectrum of the red component. We estimate the white component by subtracting the red component from the original residuals. We find the spectrum of the white component with the Z-K weighted least squares estimate. 
The next step is to fit a power-law model of the form $P_m(f) = A/(1 + (f/f_c)^2)^{\alpha/2}$ to the 
red spectral estimate for the frequency range below the frequency at which the red and white spectra cross over. We find that the ``corner frequency'' $f_c$ should not be $f_c < 1/T_{obs}$ because even if the red noise is a pure power law, fitting $\nu$ and $\dot{\nu}$ will flatten the spectrum of the residuals below this frequency.
We then compute $c(\tau)$ by Fourier transformation of $P_m(f)$ and finally obtain the covariance matrix as discussed earlier. This covariance matrix is used to re-estimate the power spectrum using the Cholesky least squares procedure. This spectrum is used to revise the model $P_m(f)$, a new $c(\tau)$ is found, a new covariance matrix, and a new Cholesky estimate of the power spectrum. At this point, the power-law model, the covariance matrix and the power spectral estimate are self-consistent. We then check to see that the whitened residuals look white, by computing their power spectrum using an OLS procedure (because they should be both white and normalized).

These steps are illustrated in Figure 4 for the pulsar J1539$-$5626. The residuals obtained from the Parkes analogue filterbank (Manchester et al. 2001)\nocite{mlc+01} are shown in the top panel (a). In the second panel (b) we show the $|DFT|^2$ power spectrum of the red component as a jagged solid line obtained by first difference pre-whitening and post-darkening. The power-law model $P_m(f)$ is shown as a smooth solid line, and the WLS spectrum of the white component is shown dotted. In the third panel (c) we show the
the Cholesky spectrum as a jagged solid line and the final model $P_m(f)$ as a smooth solid line. The original model is shown as a dashed line, and the WLS spectrum of the white component is shown exactly as in panel (b) for comparison.
One expects to see the Cholesky spectrum merge into the spectrum of the white component and this does in fact occur. Finally in the lowest panel (d) we show the OLS spectrum of $R_{W}$, the Cholesky-transformed residuals. The mean and 95\% confidence limits for a unit variance white spectrum are shown as horizontal dashed lines. One can see that the transformed residuals are in fact quite consistent with white noise.


\begin{figure}
\includegraphics[width=85mm,angle=0]{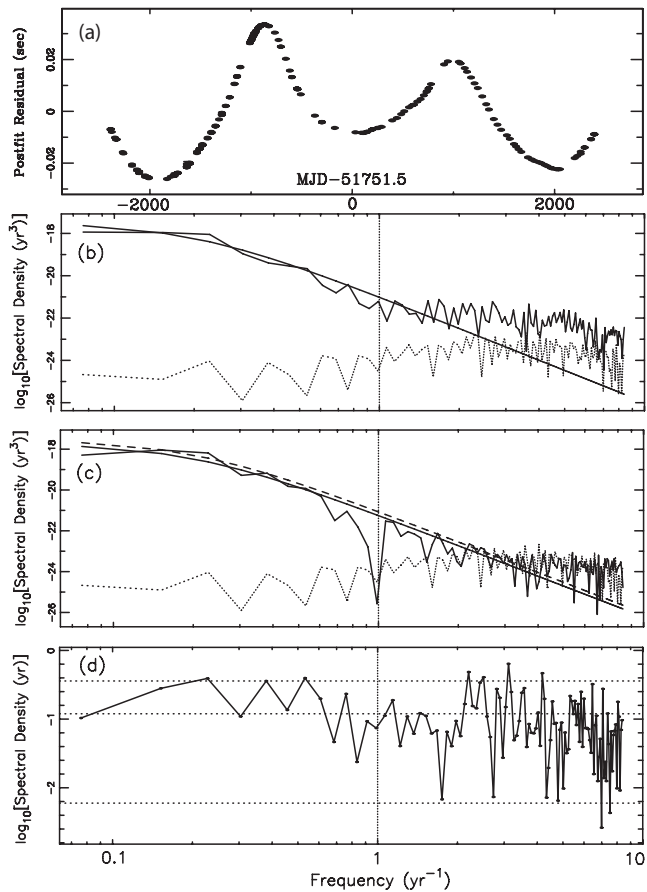}
\caption{Analysis of PSR~J1539$-$5626. Panel (a) shows the resulting residuals from a standard pulsar timing model fit for the spin-frequency and its derivative. Panel (b) shows the spectrum of the smoothed and  interpolated red component formed using the first difference pre-whitening method as a solid jagged line. The WLS  spectrum for the white component is shown dotted. The power law model $P_m(f)$ fit to the red component  is shown as a smooth solid line. Panel (c) repeats the WLS shown dotted in panel (b) and the first model $P_m(f)$  also shown as a solid line in panel (b). The Cholesky spectrum is shown as a solid (jagged) line. The revised model $P_m(f)$ fit to the Cholesky spectrum is shown dashed. Panel (d) shows the OLS spectrum of the whitened and normalised residuals with the mean and 95\%   confidence intervals shown as dotted lines.}\label{fg:1529-5626}

\end{figure}

We have provided options in {\sc Tempo2} to perform this iteration with no differencing, first-order differencing or second-order differencing. Although the differencing is only used to get a first-guess of the spectral model, we prefer to use the least order
of differencing if the spectra are similar with two different orders. For example PSR~J1539$-$5626 can be analyzed either with no differencing or first-order differencing, the results are similar. By comparison the analysis of PSR~J1939$+$2134, 
shown in Figure 1, requires
first or second-order differencing. If the exponent of the spectral model was steeper by more than a few tenths with a higher order
differencing, we would choose the higher order.  However it is always necessary to low-pass filter the residuals and interpolate them onto a regular grid. We perform the low-pass filter by convolving the residuals with a weighted exponential smoothing function
of the form $\exp(-|t/\tau_s|)$ with a time constant $\tau_s \approx 20$ days. The low pass filter response is 0.25 at 
$f = (2\pi\tau_s)^{-1}$. The smoothing time can be changed and should be adjusted so that the bandpass roughly matches the intersection of the red and white spectra. In figure 4 the intersection frequency is 1/180 days, so $\tau_s$ should be $\approx 180/2 \pi$. The results of this convolution are sampled at the original sampling times. They are then interpolated onto a regular grid using a cubic spline constrained so its step response does not overshoot (Fritsch \& Carlson 1980\nocite{fc80}). The white component is found by subtracting the red component, evaluated at the original sample times, from the original residuals.  Further details of this process are given on the {\sc Tempo2} web page.

\section{Performance Comparison}

The performance of the various fitting algorithms can be compared by simulation of observations of a pulsar with known parameters and added noise with known statistics. We expected the various algorithms to be unbiased because least-squares algorithms perform well in this respect, but we found a serious bias in the \textsc{fitwaves} algorithm which we will discuss in section 5.1.
The primary performance measure is the rms variation in the parameter estimates found by repeating the same simulation many times. It is also important that the algorithm return estimate of the uncertainties in the parameters which agree well with the actual rms variations. 

The parameters of the timing model have different effects on the residuals. Some of them are essentially time harmonic: the position and proper motion parameters adjust the amplitude and phase of an annual sine wave; the parallax does the same for a biannual sine wave; and the binary parameters adjust sine waves at harmonics of the binary period. The spin frequency parameters $\nu$ and $\dot{\nu}$ adjust the linear and quadratic polynomial coefficients and their effects are confined to frequencies $f \le 1/T_{obs}$. Jumps are Heaviside step functions having a power spectrum of the form $1/f^2$. Thus it is difficult to remove the timing noise using the polynomial or Fourier schemes without distorting the jump.
We simulated a four-dimensional test matrix: different algorithms; different sampling; different timing noise; and different parameter types. We tested four algorithms: WLS, Cholesky, polynomial, and Fourier; two sampling schemes, regular and irregular; two noise types, weak red and strong red; and three parameter types, time-harmonic, broadband, and polynomial.  The weak red noise was simulated with an amplitude of $A = 1\times10^{-24}$\,y$^{3}$, a corner frequency of $f_c = 0.3$\,y$^{-1}$ and a spectral exponent $\alpha = 2.5$.   The strong red noise had $A = 1\times10^{-17}$\,y$^3$, $f_c = 0.01$\,y$^{-1}$ and $\alpha = 5.5$.  The irregular sampling was taken from actual observations of PSR~J0711$-$6830 that contains 225 points over 14.2\,y.  The regular sampling had the same number of points sampled over the same data span. Each case was simulated 100 times with the same parameters, but different realizations of the red and white noise.  For each realization we fitted for the standard pulsar timing model parameters and for comparison recorded the  pulsar's right ascension, $\alpha$, proper motion in right ascension, $\mu_\alpha$, parallax, $\pi$, and the size of the jump.

\subsection{Bias in the Fourier method}

We found that the \textsc{fitwaves} algorithm was significantly biased, in the sense that the parameter estimates depended on the initial conditions. Parameters were biased towards zero, i.e., towards their initial conditions. So we ran special simulations to test for initial-condition bias for a harmonic parameter (proper motion in right ascension) and a broadband parameter (a phase jump). We ran 100 simulations of the same fit with slightly different initial conditions.   This showed that the WLS, Cholesky and polynomial algorithms were unbiased.  We compare the results for the Cholesky and \textsc{fitwaves} algorithms in Figure 5. In the top panel (a) the results for a phase jump are overplotted. In this case the \textsc{fitwaves} result has very small error bars but it is 100\% correlated with the initial condition. In the second and third panels (b) and (c) the results for proper motion in $\alpha$ are shown. Here one can see that the error bars for the \textsc{fitwaves} results are half those for the Cholesky results, but the \textsc{fitwaves} results are heavily biased. This bias applies to all \textsc{fitwaves} fits and therefore this technique should not be used for parameter fitting without making a careful study to confirm that it is unbiased in the application of interest. 

\begin{figure}
\includegraphics[width=90mm,angle=0]{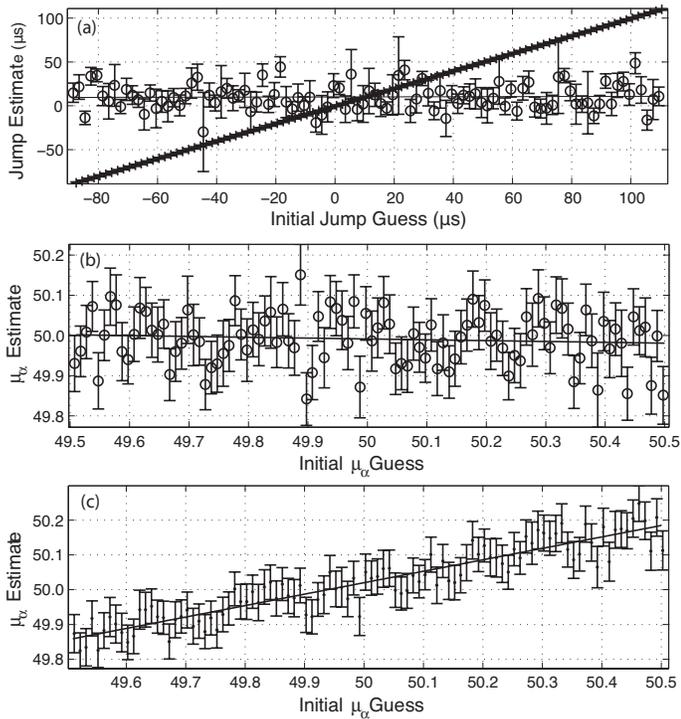}
\caption{Analysis of the bias found when using the \textsc{fitwaves}  algorithm. The parameter estimates
for a jump are shown in panel (a). Here the Cholesky points are marked  
with circles and error bars.
The \textsc{fitwaves} points are marked with error bars but the bars are so  
small that the estimates almost
appear to be a continuous diagonal line. The parameter estimates for  
proper motion in RA $\mu_\alpha$
are shown in panels (b) and (c) with the same symbols. In all cases a  
best fit straight line is drawn
through the estimates.}\label{fg:bias}
\end{figure}

\subsection{Jumps and time-harmonic parameters} 

As the WLS, polynomial and Cholesky methods were unbiased we ran the remaining simulations without changing the initial conditions. The results are shown in Table~\ref{tb:res1}.  We found that the primary measure, the rms variation, improved consistently from WLS, to polynomial, to Cholesky, as expected. The improvement was greater when there was strong red noise and irregular sampling. To facilitate comparison we have normalised the rms variation to the rms for the Cholesky method with regular sampling.  One can see that the effect of irregular sampling is large for the jump fit. This is because the jump is in a data gap, as it often is in real observations. 


\begin{table}
\caption{The ratio of the rms parameter variation to the rms parameter variation
for the regularly sampled Cholesky algorithm}\label{tb:res1}
\begin{tabular}{lllllll}
\hline
\multicolumn{4}{|c|}{\bf Regular Sampling} & \multicolumn{3}{|c|}{\bf Irregular Sampling} \\
 & WLS & Poly & Chol & WLS & Poly & Chol\\
\hline
\multicolumn{7}{|c|}{Weak red noise} \\
$\alpha$ & 1.01 & 1.02 & 1.00 & 1.81 & 1.50 & 1.28 \\
$\mu_\alpha$ & 1.20 & 1.56 & 1.00 & 1.48 & 1.26 & 0.76 \\
$\pi$ & 1.27 & 1.69 & 1.00 & 2.33 & 2.15 & 1.14 \\
Jump & 4.46 & 3.26 & 1.00 & 8.45 & 16.42 & 5.00 \\
\hline
\multicolumn{7}{|c|}{Strong red noise} \\
$\alpha$ & 17.38 & 1.53 & 1.00 & 39.74 & 3.31 & 2.96 \\
$\mu_\alpha$ & 13.32 & 1.67 & 1.00 & 28.92 & 3.24 & 2.22 \\
$\pi$ & 3.52 & 1.17 & 1.00 & 33.06 & 5.02 & 2.46 \\
Jump & 55.99 & 2.11 & 1.00 & 206.02 & 23.62 & 14.52 \\
\hline
\end{tabular}

\end{table}

In Table~\ref{tb:res2} we show the ratio of the rms of the parameter variation in the 100 realisations with the mean parameter uncertainty reported by \textsc{Tempo2}. In all cases Cholesky gave an accurate estimate of the parameter uncertainties whereas the uncertainties reported by WLS and polynomial methods were quite unreliable.


\begin{table}
\caption{The ratio of the rms parameter variation to the estimated uncertainty from the \textsc{Tempo2} fit.}\label{tb:res2}
\begin{tabular}{lllllll}
\hline
\multicolumn{4}{|c|}{\bf Regular Sampling} & \multicolumn{3}{|c|}{\bf Irregular Sampling} \\
 & WLS & Poly & Chol & WLS & Poly & Chol\\
\hline
\multicolumn{7}{|c|}{Weak red noise} \\
$\alpha$ & 1.41 & 2.70 & 1.00 & 2.37 & 3.30 & 1.24 \\
$\mu_\alpha$ & 1.52 & 3.32 & 1.05 & 2.06 & 2.76 & 1.01 \\
$\pi$ & 1.22 & 2.76 & 1.06 & 2.16 & 3.34 & 1.11 \\
Jump & 5.44 & 3.10 & 1.16 & 4.48 & 3.23 & 1.26 \\
\hline
\multicolumn{7}{|c|}{Strong red noise} \\
$\alpha$ & 0.91 & 1.96 & 0.95 & 1.45 & 1.20 & 0.90 \\
$\mu_\alpha$ & 0.83 & 2.26 & 1.14 & 1.47 & 1.53 & 0.93 \\
$\pi$ & 0.22 & 1.61 & 1.23 & 1.49 & 2.23 & 0.94 \\
Jump & 9.68 & 2.95 & 1.16 & 7.97 & 4.14 & 0.95 \\
\hline
\end{tabular}


\end{table}

\subsection{Estimation of $\nu$ and $\dot{\nu}$}

The apparent pulse frequency, $\nu$, can often be determined with ten or more decimal places.   This precision is required when predicting the pulse period and phase for observations of the pulsar, but, as the measured pulse frequency depends upon many factors including the unknown radial velocity of the pulsar, long-term drifts in terrestrial time standards and the timing noise, it does not represent the intrinsic spin frequency of the pulsar with this accuracy.  As $\nu$ and $\dot{\nu}$ are obtained from fitting a quadratic polynomial function to the timing residuals they represent the lowest frequencies ($f \le 1/T_{obs}$) in the spectrum of the residuals.  These frequencies are the most difficult to estimate because fitting the quadratic significantly modifies the residual power  $P(f \le 1/T_{obs})$.  Furthermore, at the lowest frequencies, it can be difficult to distinguish between random variations, which should be included in the whitening matrix, and deterministic variations which should be absorbed in the timing model.  For instance, ``glitch'' events during which the pulsar's rotation rate suddenly increases should be included as part of the timing model (e.g., Wang et al. 2000\nocite{wmp+00}). In Tables~\ref{tb:res3} and \ref{tb:res4} we show the results of our simulations for $\nu$ and $\dot{\nu}$.  The Cholesky method provides 3 or 4 times better parameter accuracy than a standard WLS, and the error estimates for the standard WLS are more than an order of magnitude too low. Unfortunately the Cholesky error estimates are reasonably accurate only for regularly sampled observations. For irregularly sampled observations the Cholesky error estimates are less reliable and observers will have to simulate such observations if the error estimates are important. Tools for such simulations are available in \textsc{Tempo2}.

It is clear from Figure~1 that the WLS and Cholesky methods will provide different values $\nu$ and $\dot{\nu}$ and the residuals will be different. The statistical uncertainty will be significantly smaller with the Cholesky method, but this is not the most important aspect of the analysis. If the results are to be used to interpolate the phase for comparison with other observations (for example X-ray or gamma-ray observations) then the timing noise must be modeled, interpolated, and added to the timing model. This modeling can be done with the \textsc{fitwaves} procedure. This should always be done with pulsars that have steep red timing noise, regardless of whether the fit for $\nu$ and $\dot{\nu}$ was done with the WLS or Cholesky algorithm.  Currently \textsc{Tempo2} does not include a straightforward procedure to extrapolate the phase (for instance to prepare for future observations) if the timing residuals are significantly affected by a steep red noise process. Since one has a statistical model of both the timing noise and the sampling noise, it would probably be wise to use a Wiener filter for both interpolation and extrapolation.

\begin{table}
\caption{The ratio of the rms parameter variation to the rms parameter variation
for the regularly sampled Cholesky algorithm}\label{tb:res3}
\begin{tabular}{lllll}
\hline
\multicolumn{3}{|c|}{\bf Regular Sampling} & \multicolumn{2}{|c|}{\bf Irregular Sampling} \\
    & WLS & Chol & WLS & Chol \\
\hline
\multicolumn{5}{|c|}{Weak red noise} \\
$\nu$ & 1.31 & 1.00 & 16.29 & 16.20 \\
$\dot{\nu}$ & 1.04 & 1.00 & 3.84 & 3.81 \\
\hline
\multicolumn{5}{|c|}{Strong red noise} \\
$\nu$ & 1.81 & 1.00 & 5.98 & 4.92 \\
$\dot{\nu}$ & 2.82 & 1.00 & 3.36 & 2.28\\
\hline
\end{tabular}
\end{table}

\begin{table}
\caption{The ratio of the rms parameter variation to the estimated uncertainty  from the \textsc{Tempo2} fit.}\label{tb:res4}
\begin{tabular}{lllll}
\hline
\multicolumn{3}{|c|}{\bf Regular Sampling} & \multicolumn{2}{|c|}{\bf Irregular Sampling} \\
    & WLS & Chol & WLS & Chol \\
\hline
\multicolumn{5}{|c|}{Weak red noise} \\
$\nu$ & 7.08 & 0.96 & 64.06 & 15.30 \\
$\dot{\nu}$ & 7.44 & 1.45 & 44.56 & 13.27 \\
\hline
\multicolumn{5}{|c|}{Strong red noise} \\
$\nu$ & 17.35 & 1.19 & 31.01 & 3.06 \\
$\dot{\nu}$ & 20.23 & 1.51 & 29.07 & 3.37\\
\hline
\end{tabular}
\end{table}

%


\section{Robustness to Covariance Errors}

The Cholesky method is optimal only if the covariance matrix is known, so it is important to establish its sensitivity to deviations of the covariance matrix used to obtain the whitening transformation, from the true covariance matrix of the observations. As the greatest advantage in using the Cholesky method is for steep power-law timing noise, we have simulated this case with variations in the parameters of the spectral model. The simulated spectrum had a power law exponent of $-$5.5 and a corner frequency of 0.07~${\rm y}^{-1}$. 
We adjusted the three independent parameters of the fit: the spectral exponent; the corner frequency; and the ratio of the white noise to the timing noise. We chose to use one of the harmonic parameters, the proper motion in right ascension, as the test case, with $\mu_\alpha = 10$\,mas\,y$^{-1}$. The results are shown  in Figure~6 as the observed and predicted 1$\sigma$-uncertainties versus the parameter.

\begin{figure}
\includegraphics[width=82mm,angle=0]{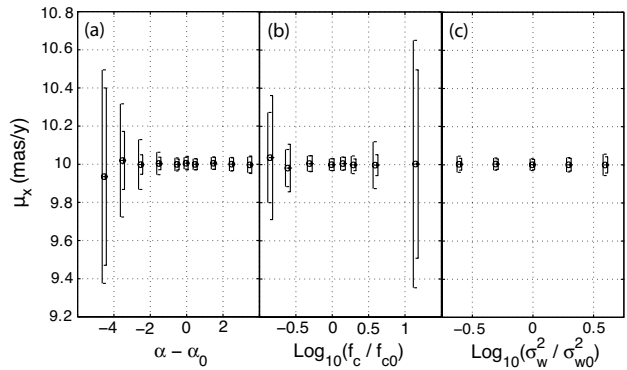}
\caption{The robustness of the Cholesky parameter estimates to errors  
in estimating the covariance
matrix of the residuals is shown using $\mu_\alpha$ as an example. The  
results come from 100 simulated
realisations of a steep red spectrum with white noise. In panel (a)  
the effect of an error in
the spectral exponent is shown. In panel (b) the effect of an error in  
the corner frequency is shown.
In panel (c) the effect of an error in the white noise level is shown.  
In each case the mean error on the parameter is shown with the right-facing error bar.  The rms of the parameter values is shown with the left-facing error bar.}\label{fg:robust}
\end{figure}

One can see that in most cases the predicted uncertainties slightly exceed the actual rms parameter variation. The cases where this difference is inverted are when the corner frequency is much less than the recommended minimum. The parameter estimate remains unbiased for all cases simulated. Generally the uncertainties are not increased significantly when the spectral exponent changes from 4 to 8; when the corner frequency is increased or decreased by a factor of two; or when the white noise variance is increased or decreased by a factor of 4. These are drastic variations which are larger than those expected in practice.

These simulations also indirectly test the effects of breakdown in the wide-sense stationarity assumption due to estimating the covariance from the post-fit timing residuals rather than the pre-fit timing residuals. The most important effect here is removing a quadratic by fitting $\nu$ and $\dot{\nu}$ which removes low frequency power and weakens the wide-sense stationarity assumption. We have tested a very wide range of variation in the low frequency power by changing the spectral exponent. The shape of the spectral model was normalized so the power was independent of the exponent at a frequency of one cycle per year. Thus power at the lowest frequencies was changed by factors of 0.01 to 100 with negligible change in the estimated proper motion. 

\section{Summary}

In the presence of red timing noise, we recommend that pulsar timing analysis always be done using the Cholesky method. Even under extreme conditions of very steep spectra, very irregular sampling, and highly variable errors it is possible to obtain an adequate estimate of the covariance function of the residuals. The Cholesky method will provide reliable error estimates except for the parameters $\nu$ and $\dot{\nu}$ and it will significantly improve the accuracy of the parameters when the residuals are very red. The Cholesky method will also be useful in combined analyses, such as using multi-frequency observations to estimate the dispersion measure, using multiple pulsars to estimate clock errors or in searching for the existence of gravitational waves, because it will properly normalise the different frequencies and different pulsars. 
The Cholesky method also provides an excellent power spectral estimate under conditions where spectral leakage would normally be a problem, i.e., steep spectra, high dynamic range, irregular sampling and variable errors. This aspect of the method has much broader application than to pulsar timing alone.

\section*{Acknowledgments} 

This work has been carried out as part of the Parkes Pulsar Timing Array project.   GH is the recipient of an Australian Research Council QEII Fellowship (project \#DP0878388), The PPTA project was initiated with support from RNM's Federation Fellowship (\#FF0348478) and JPWV is supported by the European Union under Marie Curie Intra-European Fellowship 236394.   The Parkes radio telescope is part of the Australia Telescope which is funded by the Commonwealth of Australia for operation as a National Facility managed by CSIRO.  We thank Dr. Michael Keith for helpful comments and for testing the algorithm.
  
\bibliography{journals,modrefs,psrrefs,crossrefs}

14 gid=286816
20 ctime=1310279785
20 atime=1310279786
24 SCHILY.dev=234881026
22 SCHILY.ino=2266570
18 SCHILY.nlink=1


\begin{thebibliography}{}

\bibitem[\protect\citeauthoryear{Aitken}{Aitken}{1935}]{ait35}
Aitken A.~C.,  1935, Proc. Roy. Soc., 55, 42

\bibitem[\protect\citeauthoryear{{Blackman} \& {Tukey}}{{Blackman} \&
  {Tukey}}{1959}]{bt59}
{Blackman} R.~B.,  {Tukey} J.~W.,  1959, {The measurement of power spectra,
  from the viewpoint of communications engineering. Dover Publications (New
  York)}

\bibitem[\protect\citeauthoryear{{Deller}, {Verbiest}, {Tingay} \&
  {Bailes}}{{Deller} et~al.}{2008}]{dvtb08}
{Deller} A.~T.,  {Verbiest} J.~P.~W.,  {Tingay} S.~J.,    {Bailes} M.,  2008,
  ApJ, 685, L67

\bibitem[\protect\citeauthoryear{{Edwards}, {Hobbs} \& {Manchester}}{{Edwards}
  et~al.}{2006}]{ehm06}
{Edwards} R.~T.,  {Hobbs} G.~B.,    {Manchester} R.~N.,  2006, MNRAS, 372, 1549

\bibitem[\protect\citeauthoryear{Fritsch \& Carlson}{Fritsch \&
  Carlson}{1980}]{fc80}
Fritsch Carlson 1980, J. Numer. Anal., 17, 238

\bibitem[\protect\citeauthoryear{Hobbs, Lorimer, Lyne \& Kramer}{Hobbs
  et~al.}{2005}]{hllk05}
Hobbs G.,  Lorimer D.~R.,  Lyne A.~G.,    Kramer M.,  2005, MNRAS, 360, 974

\bibitem[\protect\citeauthoryear{Hobbs, Lyne, Kramer, Martin \& Jordan}{Hobbs
  et~al.}{2004}]{hlk+04}
Hobbs G.,  Lyne A.~G.,  Kramer M.,  Martin C.~E.,    Jordan C.,  2004, mnras,
  353, 1311

\bibitem[\protect\citeauthoryear{{Hobbs}, {Edwards} \& {Manchester}}{{Hobbs}
  et~al.}{2006}]{hem06}
{Hobbs} G.~B.,  {Edwards} R.~T.,    {Manchester} R.~N.,  2006, MNRAS, 369, 655

\bibitem[\protect\citeauthoryear{{Jenkins} \& {Watts}}{{Jenkins} \&
  {Watts}}{1969}]{jw69}
{Jenkins} G.~M.,  {Watts} D.~G.,  1969, {Spectral analysis and its
  applications, London: Holden-Day}

\bibitem[\protect\citeauthoryear{Kaspi, Taylor \& Ryba}{Kaspi
  et~al.}{1994}]{ktr94}
Kaspi V.~M.,  Taylor J.~H.,    Ryba M.,  1994, ApJ, 428, 713

\bibitem[\protect\citeauthoryear{{Kramer}, {Stairs}, {Manchester},
  {McLaughlin}, {Lyne}, {Ferdman}, {Burgay}, {Lorimer}, {Possenti}, {D'Amico},
  {Sarkissian}, {Hobbs}, {Reynolds}, {Freire} \& {Camilo}}{{Kramer}
  et~al.}{2006}]{ksm+06}
{Kramer} M.,  {Stairs} I.~H.,  {Manchester} R.~N.,  {McLaughlin} M.~A.,  {Lyne}
  A.~G.,  {Ferdman} R.~D.,  {Burgay} M.,  {Lorimer} D.~R.,  {Possenti} A.,
  {D'Amico} N.,  {Sarkissian} J.~M.,  {Hobbs} G.~B.,  {Reynolds} J.~E.,
  {Freire} P.~C.~C.,    {Camilo} F.,  2006, Science, 314, 97

\bibitem[\protect\citeauthoryear{Lorimer \& Kramer}{Lorimer \&
  Kramer}{2005}]{lk05}
Lorimer D.~R.,  Kramer M.,  2005, Handbook of Pulsar Astronomy.
Cambridge University Press

\bibitem[\protect\citeauthoryear{{Lyne}, {Hobbs}, {Kramer}, {Stairs} \&
  {Stappers}}{{Lyne} et~al.}{2010}]{lhk+10}
{Lyne} A.,  {Hobbs} G.,  {Kramer} M.,  {Stairs} I.,    {Stappers} B.,  2010,
  Science, 329, 408

\bibitem[\protect\citeauthoryear{Manchester, Lyne, Camilo, Bell, Kaspi,
  D'Amico, McKay, Crawford, Stairs, Possenti, Morris \& Sheppard}{Manchester
  et~al.}{2001}]{mlc+01}
Manchester R.~N.,  Lyne A.~G.,  Camilo F.,  Bell J.~F.,  Kaspi V.~M.,  D'Amico
  N.,  McKay N. P.~F.,  Crawford F.,  Stairs I.~H.,  Possenti A.,  Morris
  D.~J.,    Sheppard D.~C.,  2001, MNRAS, 328, 17

\bibitem[\protect\citeauthoryear{{Manolakis}, {Ingle} \& {Kogon}}{{Manolakis}
  et~al.}{2005}]{mik05}
{Manolakis} D.~G.,  {Ingle} V.~K.,    {Kogon} S.~M.,  2005, {Statistical and
  adaptive signal processing: spectral estimation, signal modeling, adaptive
  filtering, and array processing}.
Artech House (Boston)

\bibitem[\protect\citeauthoryear{Petit \& Tavella}{Petit \&
  Tavella}{1996}]{pt96}
Petit G.,  Tavella P.,  1996, A\&A, 308, 290

\bibitem[\protect\citeauthoryear{{Rodin}}{{Rodin}}{2008}]{rod08}
{Rodin} A.~E.,  2008, MNRAS, 387, 1583

\bibitem[\protect\citeauthoryear{Scargle}{Scargle}{1982}]{sca82}
Scargle J.~D.,  1982, ApJ, 263, 835

\bibitem[\protect\citeauthoryear{{Thorsett}, {Arzoumanian}, {Camilo} \&
  {Lyne}}{{Thorsett} et~al.}{1999}]{tacl99}
{Thorsett} S.~E.,  {Arzoumanian} Z.,  {Camilo} F.,    {Lyne} A.~G.,  1999, ApJ,
  523, 763

\bibitem[\protect\citeauthoryear{{van Straten}}{{van Straten}}{2006}]{van06}
{van Straten} W.,  2006, ApJ, 642, 1004

\bibitem[\protect\citeauthoryear{{Verbiest}, {Bailes}, {Coles}, {Hobbs}, {van
  Straten}, {Champion}, {Jenet}, {Manchester}, {Bhat}, {Sarkissian}, {Yardley},
  {Burke-Spolaor}, {Hotan} \& {You}}{{Verbiest} et~al.}{2009}]{vbc+09}
{Verbiest} J.~P.~W.,  {Bailes} M.,  {Coles} W.~A.,  {Hobbs} G.~B.,  {van
  Straten} W.,  {Champion} D.~J.,  {Jenet} F.~A.,  {Manchester} R.~N.,  {Bhat}
  N.~D.~R.,  {Sarkissian} J.~M.,  {Yardley} D.,  {Burke-Spolaor} S.,  {Hotan}
  A.~W.,    {You} X.~P.,  2009, MNRAS, 400, 951

\bibitem[\protect\citeauthoryear{{Verbiest}, {Bailes}, {van Straten}, {Hobbs},
  {Edwards}, {Manchester}, {Bhat}, {Sarkissian}, {Jacoby} \&
  {Kulkarni}}{{Verbiest} et~al.}{2008}]{vbv+08}
{Verbiest} J.~P.~W.,  {Bailes} M.,  {van Straten} W.,  {Hobbs} G.~B.,
  {Edwards} R.~T.,  {Manchester} R.~N.,  {Bhat} N.~D.~R.,  {Sarkissian} J.~M.,
  {Jacoby} B.~A.,    {Kulkarni} S.~R.,  2008, ApJ, 679, 675

\bibitem[\protect\citeauthoryear{Wang, Manchester, Pace, Bailes, Kaspi,
  Stappers \& Lyne}{Wang et~al.}{2000}]{wmp+00}
Wang N.,  Manchester R.~N.,  Pace R.,  Bailes M.,  Kaspi V.~M.,  Stappers
  B.~W.,    Lyne A.~G.,  2000, MNRAS, 317, 843

\bibitem[\protect\citeauthoryear{{You}, {Hobbs}, {Coles}, {Manchester}, {Edwa
  rds}, {Bailes}, {Sarkissian}, {Verbiest}, {van Straten}, {Hotan}, {Ord},
  {Jenet}, {Bhat} \& {Teoh}}{{You} et~al.}{2007}]{yhc+07}
{You} X.~P.,  {Hobbs} G.,  {Coles} W.~A.,  {Manchester} R.~N.,  {Edwa rds} R.,
  {Bailes} M.,  {Sarkissian} J.,  {Verbiest} J.~P.~W.,  {van Straten} W.,
  {Hotan} A.,  {Ord} S.,  {Jenet} F.,  {Bhat} N.~D.~R.,    {Teoh} A.,  2007,
  MNRAS, 378, 493

\bibitem[\protect\citeauthoryear{{Zechmeister} \& {K{\"u}rster}}{{Zechmeister}
  \& {K{\"u}rster}}{2009}]{zk09}
{Zechmeister} M.,  {K{\"u}rster} M.,  2009, A\&A, 496, 577

\end{thebibliography}
\bibliographystyle{mn2e}

\end{document}